\documentclass[12pt]{iopart}

\expandafter\let\csname equation*\endcsname\relax
\expandafter\let\csname endequation*\endcsname\relax

\usepackage{graphicx}
\usepackage{rotating}
\usepackage{amsmath,amssymb}
\usepackage{epstopdf}
\usepackage{hyperref}
\usepackage{color}
\usepackage{cite}

\newcommand{\nn}{\nonumber}
\newcommand{\be}{\begin{equation}}
\newcommand{\ee}{\end{equation}}
\newcommand{\bear}{\begin{eqnarray}}
\newcommand{\eear}{\end{eqnarray}}

\newcommand{\md}{\mathrm{d}}

\begin{document}

\title[Nd magnon mass by spontaneous magnetization temperature dependence]
{Determination of magnon mass of neodymium magnet by temperature dependence of spontaneous magnetization}

\author{Todor M. Mishonov$^1$, Matteo Andreoni$^2$ and Albert M. Varonov$^1$}

\address{$^1$ Georgi Nadjakov Institute of Solid State Physics, Bulgarian Academy of Sciences,
72 Tzarigradsko Chaussee Blvd., BG-1784 Sofia, Bulgaria}
\address{$^2$ Anglo-American School of Sofia, 1 Siianie Strt., BG-1137 Sofia, Bulgaria}

\ead{mishonov@bgphysics.eu, S0010@aas-sofia.org, varonov@issp.bas.bg}

\date{6 November 2020, 14:54}

\begin{abstract}
The effective mass of a magnon in a neodymium magnet is evaluated using Bloch temperature dependence of spontaneous magnetization. 
We used the Hall effect magnetometer and 
thermo-couple thermometer for measuring of the temperature of the water in which the encapsulated commercial magnet is immersed.
The experiment can be repeated in every school and methodological remarks for experimental data processing are given with great detail and attention.
The purpose of the paper is to illustrate the simple relation between
the quantum spectrum of magnons 
and the easily measured magnetic field of a permanent magnet.
\end{abstract}

\date{18 Oct 2020, 20:51}

\maketitle

\section{Introduction}

The physics of magnetism is always an important part of the physics education
in high school, college and university level for undergraduate laboratories.
Let us mention some recent topics:
1) magnetic bridge sensor for observation of
superconducting-to-normal phase transition in a high-temperature superconductor and the Curie point of a ferromagnet~\cite{Kraftmakher:10},
2)  using the smartphone magnetic field sensor or magnetometer let to popularization
which can be used by every student even in home experiments~\cite{Arribas:15},
3) magnetic sensor in a smartphone as an alternate to the relatively expensive magnetic sensor probe
for accurate measurements~\cite{Taspika:18},
4) with additional efforts a compact magnetometer consisting of two Hall effect-based sensors
can be built which allows measuring the magnetic response 
of the iron oxide microparticle sample~\cite{Araujo:19}
5) mechanical set-ups require much more 	attention, but study of torque oscillations of a rotational motion
in Helmholtz coils gives a reliable method for the determination of 
magnetization of a permanent magnet\cite{Barman:19}
6)  in this direction neodymium disc magnet play important role 
to enhance knowledge of the physics of magnetism and especially to estimate 
magnetic dipole moment~\cite{Amrani:15}
7) simple, low-cost experimental setup designed for measuring the force between permanent magnets
are very appropriate for undergraduate students~\cite{Gonzalez:16}
8) magnetic measurements are very appropriate if students have complete autonomy to develop their 
experimental set-up using open-source low-cost equipment including Arduino microcontrollers and compatible sensors~\cite{Bouquet:17},
9) the theoretical studies of Ising-spin model~\cite{Guemez:91} completes the culture of education in
physics of magnetism.

Our methodological study is in the same direction.
Our purpose is to determine effective mass of magnon in the parabolic approximation of the dispersion
using cheep equipment for measuring temperature dependence of the magnetization of a neodymium magnet.
Let us return to the beginning of the quantum theory of magnetism.
Magnon is the quantum of the spin wave which have the same dependence between
the energy and momentum as a free nonrelativistic particle.
Considering microscopic Hamiltonial, Bloch\cite{Bloch:30} obtained
that parabolic dispersion of the magnon energy
\be
\varepsilon_p=p^2/2m=Ak^2,\qquad \mathbf{p}=\hbar \mathbf{k}, 
\qquad \varepsilon=\hbar\omega,\qquad A=\hbar^2/2m
\label{spectrum}
\ee
is an acceptable approximation for energies much smaller than the Curie temperature $T_c$ but 
much bigger that the frequency $\omega_f$ of the ferromagnetic resonance 
$\hbar\omega_f\ll\varepsilon\ll k_\mathrm{B}T_c$.
Later on, Landau and Lifshitz derived this spectrum via macroscopic parameters~\cite{LL9@71}. 
Roughly speaking, a magnon is a quantization of a spin-flip process, and for thermally excited magnon the  magnetization decreases with two Bohr magnetons $\mu_\mathrm{B}=q_e\hbar/2m_e=9.27\times 10^{-24}\mathrm{Am^2}$,
where $q_e$ and $m_e$ are charge and mass of the electron.
The volume density of the thermally excited magnons $n(T)$ is formally equal to
the number of thermally excited above the condensate Bose particles
\be
n(T)=\int\frac1{\exp(\varepsilon_p/k_\mathrm{B}T)-1}\frac{\mathrm{d}^3p}{(2\pi\hbar)^3}
=2\pi\times 2^{3/2} \Gamma\left(\frac32\right)\zeta\left(\frac32\right)n_q,
\quad n_q(T)\equiv\left(\frac{mk_\mathrm{B}T}{2\pi\hbar^2}\right)^{3/2}. 
\ee
For integration, spherical coordinates in momentum space were used
$\mathrm{d}^3p=\mathrm{d}\left(\frac43\pi p^3\right)$,
the dimensionless variable $x=p^2/2mk_\mathrm{B}T$,
and the well-known 
integral\cite{LL5@58}
\begin{align}
\int\limits_0^\infty\frac{x^{z-1}\mathrm{d}x}{\mathrm{e}-1}=\Gamma(z)\zeta(z),
\quad \Gamma(z)\equiv\int\limits_0^\infty t^{z-1}\mathrm{e}^{-t}\mathrm{d}t,
\quad \zeta(z)\equiv\sum\limits_{n=1}^{\infty}\frac1{n^z}.
\end{align}
In such a way we arrive at the Bloch law 
\cite{Bloch:30,LL9@71,MermAsh@33}
\begin{align}
M(T)&
=M(0)-2\mu_\mathrm{B}n(T)
=M(0)-\frac{\Gamma(3/2)\,\zeta(3/2)}{2\pi^2}
\left(\frac{2m}{\hbar^2}\right)^{\!3/2}\!\mu_{\mathrm{B}}\,(k_\mathrm{B}T)^{3/2}\\
&=M(0)-\frac{\Gamma(3/2)\,\zeta(3/2)\,\mu_{\mathrm{B}}\,(k_\mathrm{B}T)^{3/2}}
{2\pi^2A^{3/2}},\nn
\end{align}
which is a good approximation in the temperature interval 
$\hbar\omega_f\ll k_\mathrm{B}T\ll k_\mathrm{B}T_c$,
see also the well-known textbooks
\cite{Kittel@Bloch,SST@5}.
Performing linear regression in the $M$ versus $T^{3/2}$ plot
we can differentiate this linear function $M(T^{3/2})$ and its slope
\be
\left.
-\frac{\mathrm{d}M}{\mathrm{d}T^{3/2}}
\right|_\mathrm{reg}
\ee
parameterizes the effective mass of magnon
\begin{align}
m_\mathrm{eff}\equiv \frac{m}{m_e}
=\frac{\hbar^2}{2m_ek_\mathrm{B}}
\left[
{\frac{2\pi^2}{\Gamma(3/2)\zeta(3/2)}}
\left(-\frac{\mathrm{d}M}{\mu_\mathrm{B}\, \mathrm{d}T^{3/2}}
\right)
\right]^{2/3}.
\label{eq:Magn}
\end{align}
Let give some illustrative examples for determination of magnon effective mass
using suggested methods.
Usually constant $A$ from the dispersion relation Eq.~(\ref{spectrum})
is given in units eV\AA$^2$.
As
\be
\frac{\left(\hbar/\mathrm{\AA}\right)^2}{q_em_e}\approx 7620\mathrm{~meV},
\ee
where $q_e$ and $m_e$ are charge and mass of the electron,
the experimental value for Fe~(4\%~Ni) $A_\mathrm{Fe}=266\,\mathrm{meV}$\AA$^2$~\cite{Mook:73}
gives for the effective magnon mass, for methodical purposes we are giving elementary details
\be
m_\mathrm{eff,\,Fe,\,n}=\frac{m}{m_e}=\frac{\hbar^2}{2m_eA}
=\frac12\frac{\left(\hbar/\mathrm{\AA}\right)^2}{q_em_e}
\frac{q_e\mathrm{\AA}^2}{A_\mathrm{Fe}}
=\frac{7620}{2\times 266}\approx 14.3.
\label{effective_mass}
\ee
Analogously for Co~(fcc~9\%~Fe) $A_\mathrm{Co}=384\,\mathrm{meV}$\AA$^2$~\cite{Shirane:68}
gives $m_\mathrm{eff,\,Co,\,n}\approx 9.9$.
A thorough analysis of the available experimental data for the magnon mass determination will be made in a separate section of the next version of the current manuscript.


The purpose of the present work is to determine the magnon effective mass
of Neodymium magnet Nd$_2$Fe$_{14}$B$_1$
$m_\mathrm{eff,\,Nd,\,M}$
using temperature dependence of the magnetization 
using high school equipment described in the next section.
The theoretical values of the magnon mass in the case of Nd$_2$Fe$_{14}$B$_1$
and practically all ferromagnets are not yet calculated. 

\section{Experiment}
In the experiment we used disk-shaped Neodymium magnet with 
diameter $2r=19.6\,$mm and thickness $h=4.8\,$mm.
We accept $d=200\,\mu\mathrm{m}$ thickness of the protecting layer
and evaluate the volume of the magnetic material 
\be
V=\pi (r-d)^2(h-2d)=1.45 \times 10^{-6}~\mathrm{m^3}.
\ee
The magnet is in a mono-domain state and the total magnetic moment 
oriented perpendicularly to the disk is $\tilde{m}=MV$.
We will use coordinate system centered at the disc center with $z$-axis along the symmetry axes.
In such a way for the magnetic moment of the sample we have 
$\mathbf{m}=\tilde{m}\mathbf{e}_z=VM\mathbf{e}_z$.
If we measure space dependence of magnetic field $\mathbf{B}(\mathbf{R})$ at distances $R>>r$
we can use the formula for the field of a point magnetic moment\cite{Feynman@MM,LL2@MM,Allah}
\begin{align}&
\mathbf{B}=\frac{\mu_0}{4\pi}
\frac{3\mathbf{n}(\mathbf{n}\cdot\mathbf{m})-\mathbf{m}}{R^3},\qquad
\mathbf{n}\equiv\frac{\mathbf{R}}{R},\qquad
\frac{\mu_0}{4\pi}=10^{-7} \, \mathrm{T m/A}.
\end{align}
In the Gaussian system, $\mu_0/4\pi=1$ and we use system invariant form.
Let us mention two special cases for the component of the magnetic field parallel 
to the symmetry axis $B_z=\mathbf{B}\cdot\mathbf{e}_z$
\begin{align}&
B_z=\frac{\mu_0}{4\pi}\frac{2\tilde{m}}{R^3},\qquad 
M=\frac12\frac{4\pi}{\mu_0}\frac{B_zR^3}{V},\qquad
\mbox{for}\quad \mathbf{n}=\mathbf{e}_z,\label{e_z}\\&
B_z=-\frac{\mu_0}{4\pi}\frac{\tilde{m}}{R^3},\qquad
M(T)=-\frac{4\pi}{\mu_0}\frac{B_z(T) R^3}{V},\qquad
\mbox{for}\quad \mathbf{n}=\mathbf{e}_x.
\label{e_x}
\end{align}
These approximative equations above are applicable when assistance from magnet  to observation point $R$ is much bigger than the radius $r$ of the disk shaped magnet $R>r$ 
and the misalignment of the experimentally measured distance.

\begin{figure}[h]   %
\centering
\includegraphics[scale=0.05, angle=270]{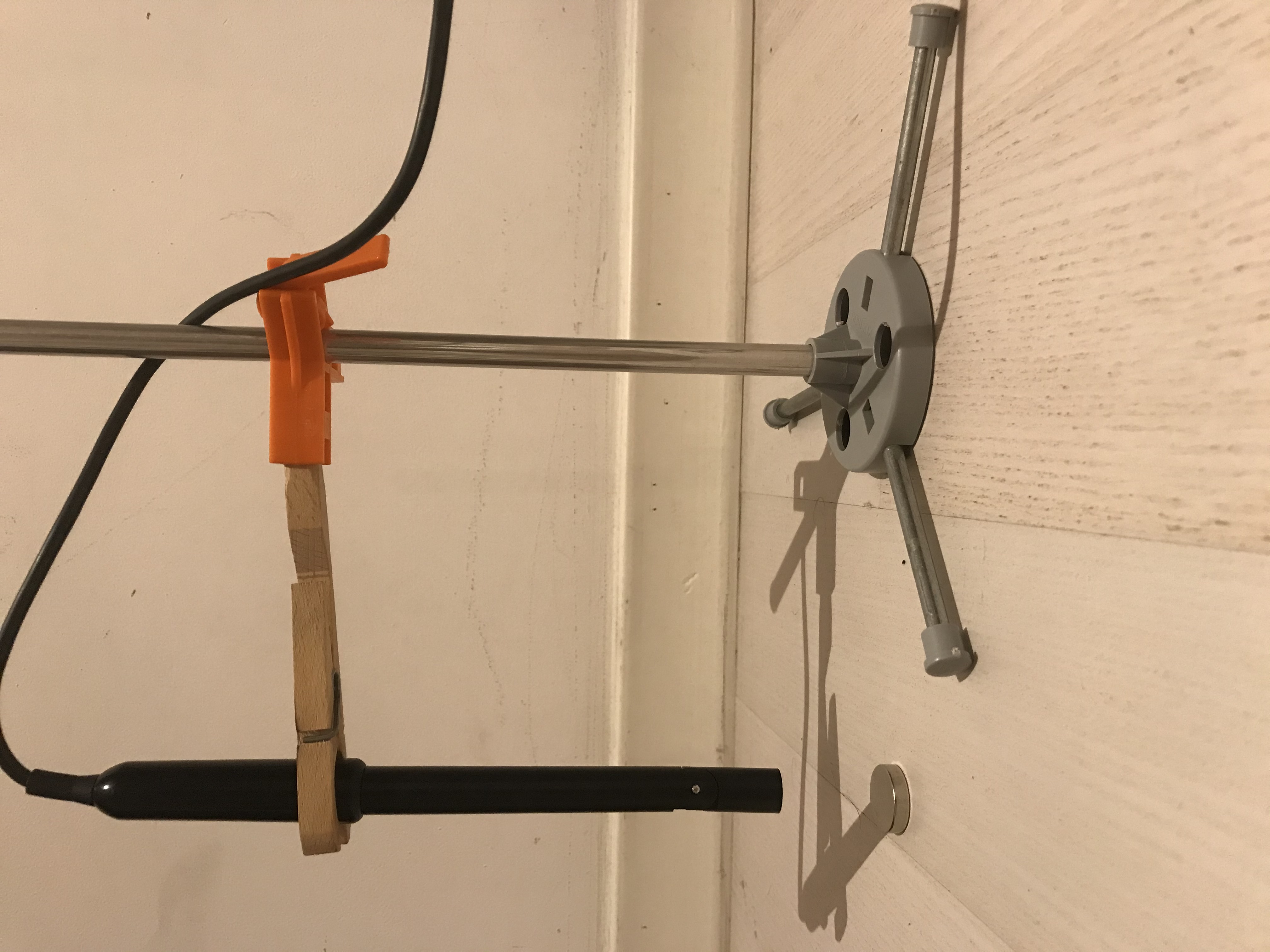}
\caption{Set up for the measurement of the magnetic field along the axis of symmetry of the magnet.
The magnetic field sensor is fixed by the holder.}
\label{fig:b-rsetup}
\end{figure}

\begin{figure}[h]      %
\centering
\includegraphics[scale=1.4]{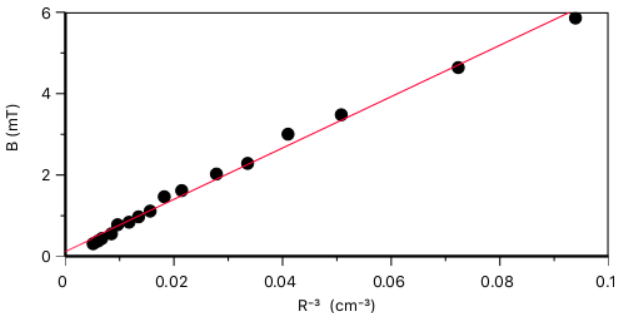}
\caption{Magnetic field $B_z$ versus distance $R^{-3}$ along the symmetry axis
$\mathbf{n}=\mathbf{e}_z$.
The slope of the linear regression $\md B/\md R^{-3}=63.09\times10^{-9}\,\mathrm{T/m^{-3}}$
According Eq.~(\ref{e_z}) the slope of the regression gives
magnetic moment $\tilde{m}=0.315\, \mathrm{Am^2}$ @ $25^\circ$~C
for the used neodymium magnet.
The interval of the distances is $22\,\mathrm{mm}<R<58\,\mathrm{mm}$.
The misalignement or a change in height of 1 mm has 
negligible influence on the measured values. 
}
\label{fig:b-r}
\end{figure}

\begin{figure}[h]   %
\centering
\includegraphics[scale=0.05]{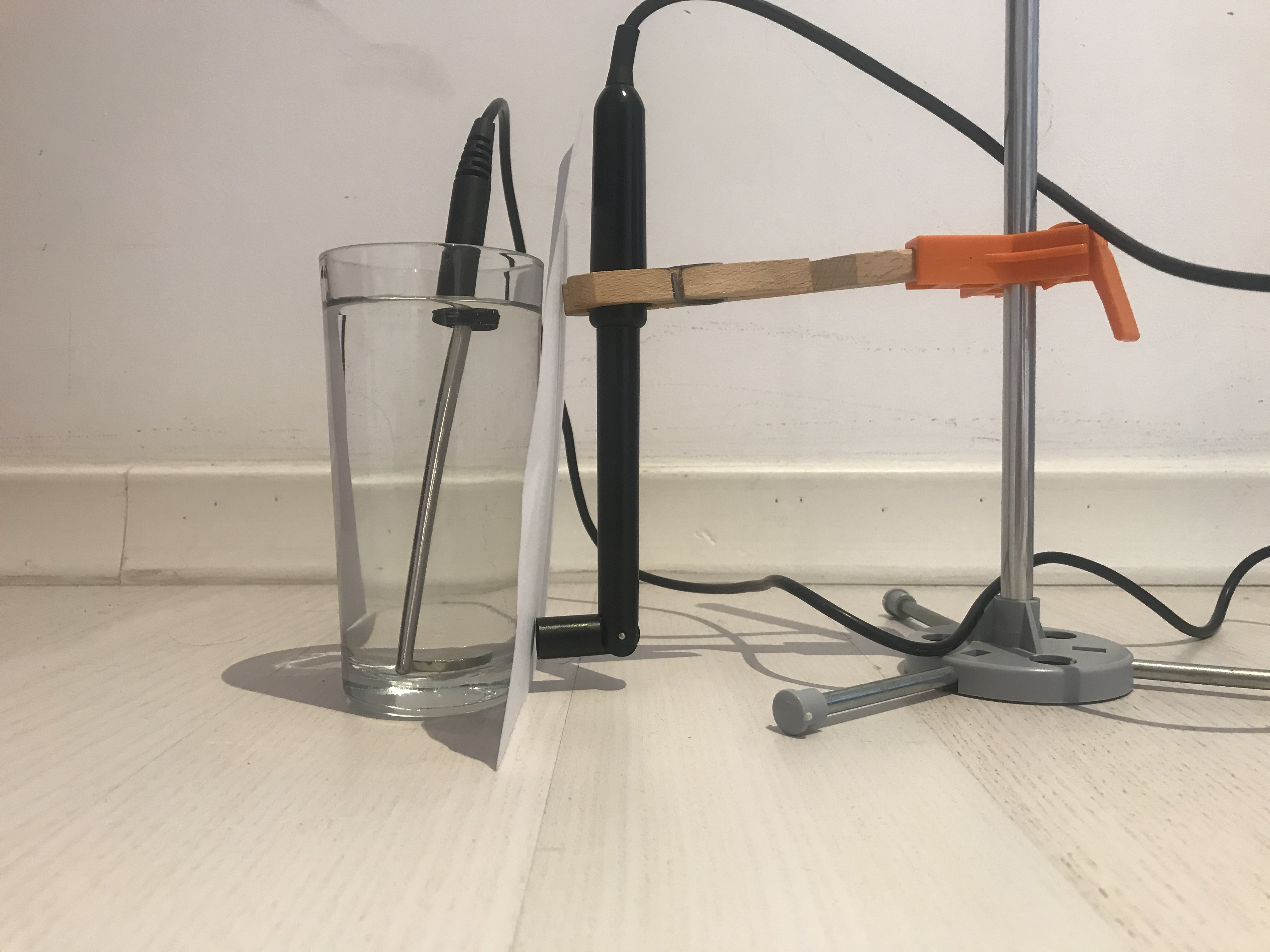}
\caption{Setup for the temperature dependence the magnetization $M(T)$ according Eq.~(\ref{e_x}).
The disk shaped permanent magnet is on the bottom of a glass cup with boiling water.
The thermo-couple Vernier$^\copyright$ temperature probe is also immersed in the water.
And the magnetic sensor is out of the cup $\mathbf{n}=\mathbf{e}_x$ at a fixed distance $R=2.5$~cm.
Simultaneously time dependence of the temperature $T(t)$ and magnetic field $B_z(t)$ are measured.
The magnetic field strength and temperature of the magnet was recorded simultaneously every 15 seconds as it cooled down to room temperature. 
The uncertainties of the magnetic field sensor and temperature probe were
$\pm 0.35$~K and $\pm 0.004$~mT, respectively \cite{MermAsh@33,Holtzberg:64}. 
In such a way is determined the temperature dependence of the magnetic field $B_z(T)$ drown in 
Fig.~\ref{fig:b-T}.
}
\label{fig:b-Tsetup}
\end{figure}

\begin{figure}[h]    %
\centering
\includegraphics[scale=0.5]{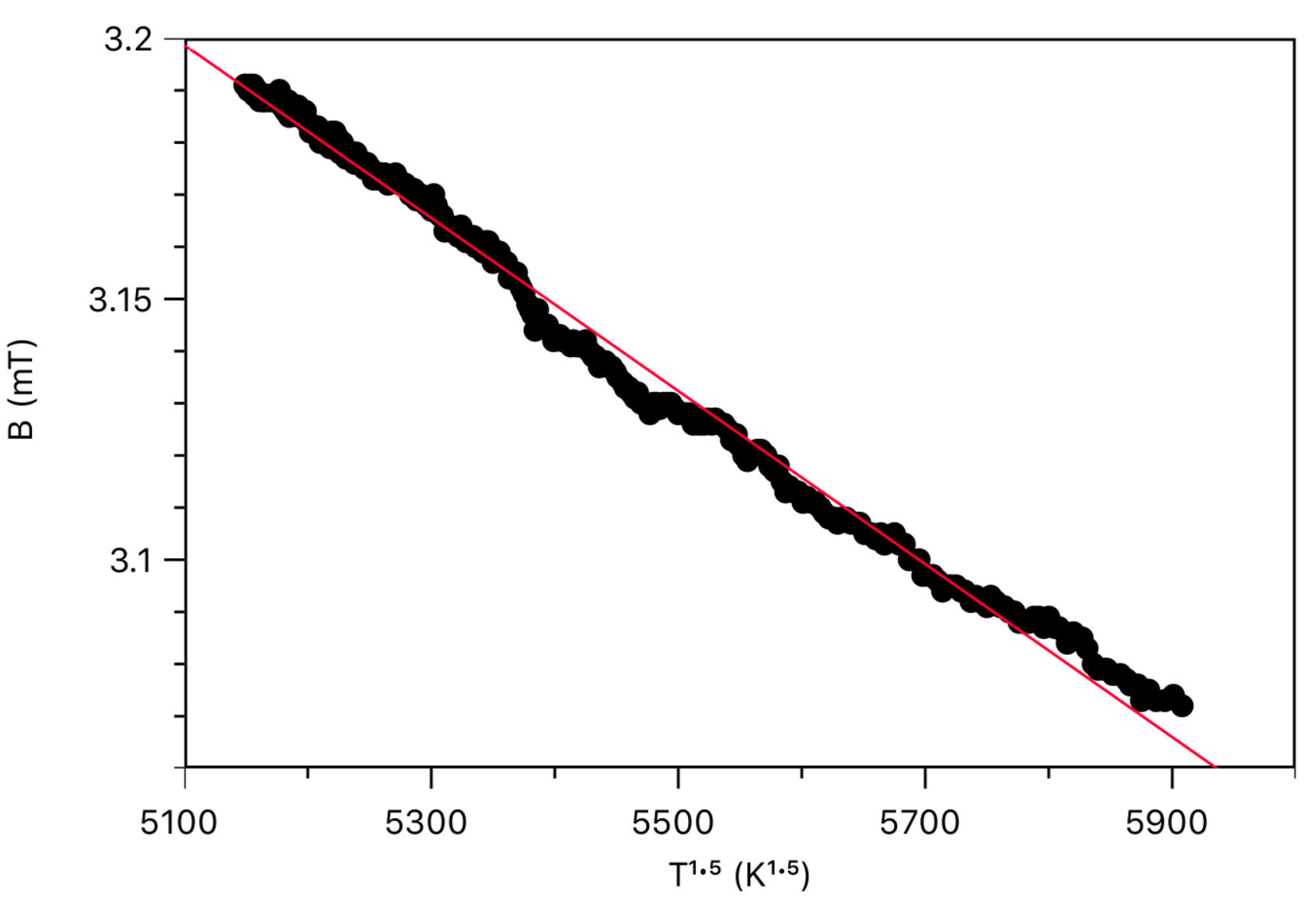}
\caption{Magnetic field $B_z$ versus temperature $T^{1.5}$ (Bloch plot)
for $\mathbf{n}=\mathbf{e}_x$ geometry and fixed distance $R$.
According to Eq.~(\ref{e_x}) the used volume of the magnet $V$
gives the temperature dependence of the volume density of magnetization $M(T)$
used for determination of effective mass of neodymium magnet
according to Eq.~(\ref{eq:Magn}). 
The slope of the linear regression $\md B/\md T^{3/2}=\!-166\, \mathrm{nT/K^{3/2}}$.
Temperature interval of this measurement is 
$25.03^{\circ}\mathrm{C}<T<70.71^{\circ}\mathrm{C}$.
}
\label{fig:b-T}
\end{figure}

In Fig.~\ref{fig:b-rsetup} the set-up for measuring of the magnetic filed $B_z$
along the symmetry axis with observation point $\mathbf{R}=R\mathbf{e}_z$ is depicted. A Vernier$^\copyright$ Magnetic Field Sensor was used for all magnetic field measurements.
The result of the experiment is shown in Fig.~\ref{fig:b-r}.
After this recall of the well-known formula for the magnetic dipole
we can address the study of the temperature dependence of the spontaneous
magnetic moment $M(T)$.
The setup can be seen in Fig.~\ref{fig:b-Tsetup}. In this experiment, both the Vernier$^\copyright$ Magnetic Field Sensor and the Vernier$^\copyright$ Temperature Probe were used to measure the magnetic field strength and temperature of the magnet, respectively. The neodymium magnet was placed in a glass of boiling water together with a Vernier$^\copyright$ Temperature Probe. 
The  Vernier$^\copyright$ Magnetic Field Sensor was set up 1~cm from the edge of the glass, and 2.5~cm from the magnet, and a sheet of paper was placed between the glass and the sensor to act as a heat barrier, preventing any heat irradiated by the glass to affect the readings of the magnetic field sensor.  
The uncertainties of the Magnetic Field Sensor and Temperature Probe were
$\pm 0.35$~K and $\pm 0.004$~mT, respectively \cite{MermAsh@33,Holtzberg:64}. 
The experimental data obtained by this set-up are represented in Fig.~\ref{fig:b-T}.
The experimental data processing of the measurements depicted in Fig.~\ref{fig:b-Tsetup}
according to Eq.~(\ref{eq:Magn}) 
gives effective mass of Nd magnet
$m_\mathrm{eff,\,Nd}= 30$ which is the goal of the present study.
We have obtained some rough evaluation of the magnon mass parameterizing Eq.~(\ref{spectrum}).
It will be very interesting to compare this result with reliable neutron scattering measurements
in order to check the accuracy of determination of spectrum parameters 
by high-school thermodynamic measurements.

\section{Discussion and conclusion}

As far as we know, such a methodical experiment is not present in the usual university laboratory exercises worldwide,
meaning that this is novel and original laboratory experiment based on relatively inexpensive equipment.
One of the authors of the current paper MA was a high school student during this experimental research, 
he successfully performed the measurements, the experimental data processing, and understood the theory.
With motivated and dedicated mentors,


\section*{References}
\bibliography{refs}

\end{document}